# A PRINCIPLES-BASED ETHICS ASSURANCE ARGUMENT PATTERN FOR AI AND AUTONOMOUS SYSTEMS

**Zoe Porter, Ibrahim Habli, John McDermid, Marten Kaas**

## ABSTRACT

An assurance case is a structured argument, typically produced by safety engineers, to communicate confidence that a critical or complex system, such as an aircraft, will be *acceptably safe* within its intended context. Assurance cases often inform third party approval of a system. One emerging proposition within the trustworthy AI and autonomous systems (AI/AS) research community is to use assurance cases to instil justified confidence that specific AI/AS will be *ethically acceptable* when operational in well-defined contexts. This paper substantially develops the proposition and makes it concrete. It brings together the assurance case methodology with a set of ethical principles to structure a principles-based ethics assurance argument pattern. The principles are justice, beneficence, non-maleficence, and respect for human autonomy, with the principle of transparency playing a supporting role. The argument pattern – shortened to the acronym PRAISE – is described. The objective of the proposed PRAISE argument pattern is to provide a reusable template for individual ethics assurance cases, by which engineers, developers, operators, or regulators could justify, communicate, or challenge a claim about the overall ethical acceptability of the use of a specific AI/AS in a given socio-technical context. We apply the pattern to the hypothetical use case of an autonomous 'robo-taxi' service in a city centre.

## 1. INTRODUCTION

Artificial Intelligence (AI) is one of the most significant technological developments of our times and its use is increasingly pervasive.[1] Whether in AI-enabled recommender or decision-support systems, or in autonomous systems (AS) which influence the environment with greater degrees of independence from direct human intervention and control, AI is being integrated into the operations of virtually every conceivable sector: agriculture; automotive; aviation; criminal justice; defence; education; energy; finance; healthcare; the humanitarian sector; insurance; manufacturing; maritime; nuclear; the police; retail; the sciences (physical, life, and earth); social care; space [3-5]. The raft of consumer applications is also growing, including home safety, consumer imaging systems, and personal monitoring [6, 7]. In addition, AI is ubiquitous across the internet and embedded in online services, whether virtual assistants, immersive maps, or personalised search. AI-generated content utilising large language models (LLMs) now portends a new transformative wave of the technology [8].

---

[1] We take the broadly functionalist view that AI refers to a set of computational techniques which enable machines to do what it takes intelligence for humans to do. This encompasses a range of techniques including data-driven *machine learning* (ML) and *logic and knowledge-based approaches* [1]. Although defining AI in this way covers many systems that are now considered 'traditional', we adopt this definition in order to take a broad view of AI, rather than identify it with any single technique. As noted in the OECD's definition, AI systems *"are capable of influencing the environment by producing an output (prediction, recommendation or decision) for a given set of objectives ... [and] are designed to operate with varying levels of autonomy."* [2].

Over the past five to ten years, concerns about the ethical impact of these technologies have led *"seemingly every organisation with a connection to technology policy … [to] author or endorse a set of ethical principles for AI/AS"* [9]. Notable examples at the international and governmental level include: the Asilomar Principles in 2017 [10]; the Montréal Declaration for Responsible AI in 2018 [11]; the UK House of Lords Select Committee report on AI in 2018 [12]; European Commission High-Level Expert Group (HLEG) on AI in 2019 [13]; the OECD AI Principles in 2019 [14]; the Beijing AI Principles in 2019 [15]; and UNESCO's Recommendation on the Ethics of Artificial Intelligence in 2022 [16].

Broadly, the range of research issues concerning the ethics of AI can be divided into three categories. First, the ethics of data-driven machine learning (ML). ML-based systems are trained on large datasets to perform classification and regression tasks [17]. The use of data-driven ML incurs several ethical hazards, including: the codification of stereotypes and historical bias in training datasets against specific demographic groups [18-21]; the opacity of extremely complex ML models, such as deep neural networks (DNNs) [17]; and the use and misuse of personal data [22, 23]. Second, there are novel issues raised by autonomous systems (AS) which replace human decision-makers who have historically been relied on as the primary form of hazard mitigation [24], as well as the locus of judgement [25]. These include: system safety [24]; the impact of increasingly autonomous machines on human autonomy [26]; and the location of moral accountability and legal liability for accidents and adverse consequences [27, 28]. Third, shifts in risk distribution prompt ethical questions, for example when, despite overall benefits, risk is disproportionately weighted with stakeholder subgroups, such as vulnerable users [29]. We might add to this category questions about the long-term environmental impact of AI development and use [30, 31].

In the past couple of years, ethical principles for AI have started to translate into proposals for regulation, such as the European Commission's draft proposal for an Artificial Intelligence Act [32] and the U.S. proposal for an Accountability for Algorithms Act 2022 [33], and the UK's recent white paper outlining a regulatory framework for AI [34]. But much of what is written at the governmental level is abstract and paths to implementation remain unclear. Various tools and techniques for trustworthy AI supplement these proposals, including: technical standards; algorithmic auditing; model cards; privacy-preserving techniques; fairness metrics; red-teaming; ethical black boxes; impact assessments; performance testing; and conformity assessment [35-42]. But this work typically considers ethical desiderata as discrete requirements, and frameworks to reason about tensions and trade-offs between ethical goals and values remain necessary [43-45].

The proposal advanced in this paper addresses the following research problem. Given the range ethical concerns about AI/AS, how can we construct a practical framework that supports reasoning about the overall ethical acceptability of the use of systems in a joined up, or holistic, way and which takes into consideration their specific socio-technical contexts?

To address the research problem, we bring together the assurance case methodology, most often used by engineers to communicate that a system is safe for use in a given operational context, with a set of ethical principles, in order to construct a framework for communicating justified confidence (or challenging confidence) in the overall ethical acceptability of the use of a system in its intended context.

More specifically, the paper presents and describes an ethics assurance argument pattern – a general template on which individual ethics assurance cases for specific AI/AS in well-defined contexts could be based. The argument pattern is structured around goals which correspond to the following four core ethical principles: justice; beneficence; non-maleficence; and respect for human autonomy. The ethical principle of transparency plays a supporting role within the framework. Together, these are referred to as the '4+1 ethical principles'.[2] The ensuring argument pattern enables a wide set of ethical concerns and desiderata for specific AI/AS to be reasoned over within a single framework.

The definition of 'ethically acceptable' employed in the proposed framework is grounded in the notion of a social contract. That is, it is based in the idea that the use of a given AI/AS would be ethically acceptable if none of the affected stakeholders could reasonably object to its use in the intended context. Underlying this, in turn, is an equal respect for all affected stakeholder groups and a commitment to ensuring there is an equitable distribution of benefit, tolerable residual risk, and respect for human autonomy across them. The framework proposed is therefore more ambitious and sets a higher threshold for 'ethical acceptability' than much proposed regulation – although it is consistent with emerging regulatory frameworks in covering many of the same ethical desiderata. The acronym PRAISE (PRinciples-bAsed EthIcs

---

[2] This to some extent mirrors the '4+1' principles of software safety assurance, although those are technical and not ethical principles [46].

assurance) seems apt in view of this ambitious approach.

To be clear, what is presented is *not* an outline reasoning process to be formulated as algorithms and automated by a computer. Nor is it a boilerplate to be worked through *mechanically* by humans. It is, rather, a framework to structure and support human deliberation and judgement. Used in this way, the aim is that the PRAISE framework can help human decision-makers to justify, communicate, and challenge confidence that it would be ethically acceptable to deploy and use a specific system in a specific real-world context.

The paper introduces the reader to the assurance case methodology and the 4+1 ethical principles, then presents their combination in the PRAISE argument pattern and describes this at a relatively high level of abstraction. It proceeds as follows.

**Section 2** introduces the assurance case methodology. Assurance cases and argument patterns can be presented in different notations. The paper adopts the widely used, graphical Goal Structuring Notation (GSN). The core elements of GSN are introduced.

**Section 3** describes the ethical principles which are translated into key sub-goals in the argument pattern and explains the rationale for using these to structure the framework. The four core principles are: justice; beneficence (do good); non-maleficence (do no unjustified harm); and respect for human autonomy. They are supported by a principle of transparency, which refers to transparency of the assurance process as well as transparency of the AI/AS, and is the '+1', giving the '4+1' ethical principles.

**Section 4** provides an overview of the argument pattern and its logical flow. In **Section 5**, the PRAISE argument pattern is described in more detail. Sections 4 and 5 are illustrated with the hypothetical use case of an autonomous 'robo-taxi' service in a city centre.

**Section 6** reiterates how the proposed PRAISE argument pattern addresses the research problem, and identifies the next steps for advancing, refining, and evaluating the methodology with real-world use cases.

## 2.  THE ASSURANCE CASE METHODOLOGY

'Assurance' refers to the general activity of providing justified or warranted confidence in a property of interest. Within engineering, this property is, most commonly, safety [47]. The proposed framework takes a broader property of interest: ethical acceptability.

## 2.1.  The Assurance Case Methodology

The assurance case methodology is one of several methods safety engineers use to manage risk from complex and critical systems. Specifically for safety, it is a methodology for presenting *"a clear, comprehensive and defensible argument that a system is acceptably safe to operate within a particular context"* [48: 3].

When we speak of an 'assurance argument pattern', we mean a general, reusable template for reasoning about a particular issue at an abstract level [49]. When we speak of an 'assurance case', we mean an instantiated, auditable, and compelling argument that a given system (or service or organisation) will operate as intended for a defined application in a defined environment. Assurance cases are supported by a body of evidence and assumptions are made explicit [50, 51]. Assurance cases are often based on the template of an argument pattern. The proposal is that the principles-based ethics assurance argument pattern presented in this paper – the PRAISE argument pattern – provides a template for individual ethics assurance cases. The intent is also to stimulate debate around ethics assurance cases as a promising methodology to achieve and assure ethically acceptable uses of AI/AS.

Assurance cases are often required as part of the regulatory process. In highly regulated sectors, such as defence, aviation, and healthcare, they tend to inform pre-deployment approval of a system as safe to operate. Different drivers have led to the adoption of assurance and safety cases in industry [52]. High profile accidents are one key driver. For example, following the Piper Alpha oil platform explosion in 1988, the public inquiry chaired by Lord Cullen led to the formal adoption of safety cases in the UK offshore industry [53]. Another key driver is system complexity and its potential impact on safety management. This is perhaps best illustrated by the requirement for safety cases by the automotive standard ISO 26262 [54], prompted by the increased technical complexity of the embedded electronics as well as the organisational complexity of the supply chain.

Assurance cases can be presented using different notations. One standard notation is the Goal Structuring Notation (GSN), which was developed at the University of York in the 1990s [48, 55]. The PRAISE argument pattern described in Sections 4 and 5 uses GSN. This notation is selected because of its widespread use in industry, the existence of a detailed standard [50], and the explicit support it provides for argument patterns [49].

GSN is based on a model of informal argumentation developed by the philosopher Stephen Toulmin [56] and it places an emphasis on providing a well-structured justification or warrant for key claims. Assurance cases structured in accordance with GSN are hierarchically decomposed. They argue from the top-level goal – the key claim that the argument supports – typically via an argument strategy which spells out the inference between this goal and the sub-goals that support it. These sub-goals in turn are supported by solutions (i.e., references to evidence items), perhaps after several further levels of decomposition. Transparency of the argument enables confidence in each sub-goal to be evaluated. The argument strategy, sub-goals, and their supporting evidence together provide abductive support for the claim expressed in the top-level goal. As Goodenough and colleagues put it, *"an assurance case provides defeasible reasons for believing that a claim is true"* [57: 27].

The main symbols and elements of a GSN assurance argument are presented in the legend in Figure 1. Many of these key elements are contained in the PRAISE argument pattern described in Sections 4 and 5.

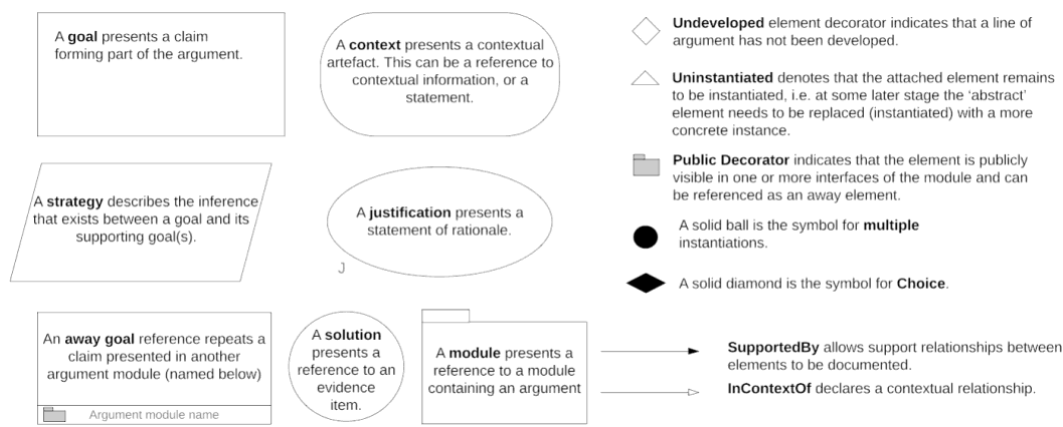

**Figure 1**
**Symbols and elements of a GSN argument**
**Extracted and adapted from Assurance Case Working Group [50]**

## 2.2.    Beyond Safety Assurance: Extending the Assurance Case Methodology to Assure the Ethical Acceptability of AI/AS

The idea of developing the assurance case methodology beyond safety to justify confidence in claims that AI/AS will be ethically acceptable is a growing proposition amongst AI ethicists and researchers focused on developing actionable policies and procedures [58-63]. This paper substantially develops that proposition and makes it concrete.

The assurance case methodology has several distinctive features [64, 65] that lend well to reasoning about the ethical acceptability of AI/AS [59]. Its explicitness enables scrutiny, discussion, debate, and continued improvement, which is important if the use of the systems is going to be ethically acceptable over the long term. Assurance cases also provide a framework for integrating and consolidating multiple evidence sources – from dataset audits to impact assessments – that will likely be required to substantiate claims about ethical AI/AS [36, 39], as well as to comply with emerging regulation and law [32-34]. Furthermore, because they are clear and accessible to the observer without specialist knowledge, assurance cases can foster interdisciplinary and multidisciplinary collaboration.[3] This is key to effective solutions in the complex arena of ethical AI/AS.

There is already a healthy pluralism in the new field. Some approaches use the assurance case methodology to address discrete ethical properties, such as fairness or explainability, in separate ethical assurance argument patterns [60, 61]. Other approaches attempt to address a range of ethical desiderata within a single argument pattern [62]. In this paper, we take the latter approach. The aim is to cover a heterogeneous collection of ethical considerations that need to be addressed together to justify claims of *overall* ethical acceptability from the use of an AI/AS.

There is also emerging variation in what we call 'top-down' and 'bottom-up' approaches to ethical assurance. Top-down approaches specify the structure and decomposition of the assurance argument [60-62]. In bottom-up approaches, directly affected stakeholders actively engage in the design of the argument pattern [58, 59], in line with the Responsible Research and Innovation policy framework [67].

The approach taken in this paper is a 'hybrid' one, combining top-down and bottom-up elements.[4] We 'pre-structure' the argument pattern and provide a detailed template. This is the 'top-down' element. But the strong intent is also that directly affected stakeholders will participate in the instantiation and validation of ethics assurance cases based on this template, for specific use cases. This is the 'bottom-up' element of the process. This participatory instantiation and validation may lead to revising the argument pattern overall or adapting it to specific use cases.

## 3. THE 4+1 ETHICAL PRINCIPLES

The PRAISE argument pattern is structured around four core ethical principles: justice;

---

[3] To clarify, we take 'interdisciplinary' to mean the synthesis of different disciplinary perspectives in the final research output, and 'multidisciplinary' to mean the drawing upon different disciplines that still stay within their boundaries in the research output [66]. As an example, this paper puts forward an interdisciplinary proposal.

[4] [58] also take a hybrid approach, but with a greater bottom-up emphasis since the decomposition of principles is determined by the directly affected stakeholders.

beneficence; non-maleficence; and respect for human autonomy. Each of these four principles are translated into sub-goals – key claims that the argument supports.  Transparency is not treated as a core ethical principle in the argument but as a supporting principle (the '+1' of the '4+1').

## 3.1.  Ethical Principles for AI/AS

Between 2014 and 2019, well over 80 major sets of ethical principles for AI/AS were published by public and private sector bodies [68]. Meta-analyses of these documents have revealed consensus around the key ethical values of concern, such as transparency, justice, non-maleficence, responsibility, privacy, beneficence, freedom and autonomy, trust, sustainability, dignity, and solidarity [9, 68].[5] This widespread endorsement of similar values establishes a starting point.

We agree with, and build on the insights of, a position already expressed [44, 69, 70] that there is a striking overlap between these values endorsed for ethical AI/AS and the four classical principles of biomedical ethics [71, 72].[6]  The overlap has been recognised by the OECD, amongst others [14, 73].

The ethics assurance argument pattern presented in this paper is structured around the following four core ethical principles, which are most closely associated with biomedical ethics [71, 72] but can be adapted to the context of ethical AI/AS:

- justice (the distribution of benefits and risks from use of the system should be equitable across affected stakeholders);
- beneficence (the use of the system should benefit affected stakeholders);
- non-maleficence (the use of the system should not cause unjustified harm to affected stakeholders);
- respect for human autonomy (affected stakeholders' capacity to live and act according to their own reasons and motives should be respected).

The usability of the four principles in the medical domain [74] suggests that they could be a good heuristic for a wide range of non-philosophical professionals, including designers, engineers, safety teams, manufacturers, operators, and users.

---

[5] Ethical principles can be understood as ethical values (e.g., the value of fairness) expressed in normative form (e.g., the principle that public institutions should uphold fairness).
[6] Noting that the biomedical principles concern *personal* autonomy, to avoid risk of undue paternalistic medical interventions, rather than *human* autonomy in the face of increasingly autonomous software-based systems.

By using these four principles to structure the PRAISE argument pattern we can also capture a range of ethical desiderata in a single framework. The summary in Table 1 shows that the many of the ethical values endorsed for AI/AS can be covered by these four principles.

| | Beneficence | Non-maleficence | Human Autonomy | Justice |
|---|---|---|---|---|
| *Principle Covers* | Well-being Flourishing Sustainability Common good | Well-being Safety Security Privacy Non-discrimination | Human control Self-determination Consent Dignity Empowerment Liberty | Fairness Equity Non-discrimination Reciprocity |

**Table 1**

**Coverage of the four core ethical principles**

To be clear, not all the proposed values fit neatly under one core principle alone. Moreover, not all of those that have been proposed in the ethics and policy debate are clearly derivative of the proposed four. 'Transparency' is one such, but it is included in the PRAISE argument pattern in a supporting role: it is the '+1' to the four core ethical principles.[7] 'Trust' is another example. But, plausibly, the enactment of the '4+1' ethical principles would provide good grounds for trust. The same might be said for 'solidarity' – that unity and cohesion would arise from the enactment of the principles, and particularly the principle of justice. [8]

Let us now set out the '4+1' ethical principles as they are interpreted in this paper. This sets the scene for the discussion of the PRAISE argument pattern in Sections 4 and 5.

The *principle of justice*, as we construe it, requires that the balance of benefits, tolerable residual risks, and constraints on human autonomy from the use of an AI/AS is equitable across affected stakeholders. This is a principle of distributive justice. The equitable distribution is understood in terms of *who* benefits from use of the AI/AS and at what cost to others (and themselves) and *who* bears the residual risks and at what benefit to themselves (and others). This reflects the 'social contract' approach to ethical acceptability that is integral to the PRAISE framework. The guiding idea is that, if this distribution is equitable across affected stakeholders, hypothetical rational agreement across affected stakeholders would be reached. This agreement, in turn, is what makes the use of the AI/AS ethically acceptable. As shown in the graphical overview of the PRAISE argument pattern in Section 4, the principle of justice takes priority over the other ethical principles in the structure of the framework.

---

[7] In the 4+1 safety principles [46] was confidence, with the understanding that greater risk implied the need for greater confidence in the argument and evidence. In future developments of the PRAISE approach, we intend to address whether the level of transparency should increase with the level of risk, constraints to human autonomy, or disparities in the balance of benefit, risk, and constraints on human autonomy across affected stakeholders.

[8] Another exclusion is 'responsibility'. We do not include a principle of responsibility in the PRAISE argument pattern. It is a complex normative goal involving distinct methodologies, particularly in law, that is beyond the scope of this paper to address here. We envisage that a separate responsibility assurance argument would ultimately connect this argument pattern.

The *principle of beneficence* requires that affected stakeholders benefit from the use of the AI/AS. We seek to reinforce that "*there should be a (sought-after) benefit of having the system in the first place*" [30: 4]. Typically, the rationale given for the development and deployment of AI/AS is the huge benefits to human welfare that these systems will unlock, not just in terms of driving economic growth, but also scientific breakthroughs, streamlined systems, improving mobility, or relieving humans of dangerous and dirty tasks [34, 75-77]. The inclusion of the principle of beneficence aims to make these aims concrete. It is notable that the notion that AI/AS should be beneficial for humanity is central to some of the sets of principles outlined above [10, 11, 13-15]. But this does not seem to be being actively translated into regulation and public policy.

The *principle of non-maleficence* requires that the use of an AI/AS does not cause unjustified harm. The risk of physical harm is the traditional arena of safety engineering and safety assurance. Safety methods, such as Hazard and Operability Studies (HAZOP) [78], and standards, such as ISO 21448 [79] which covers Safety of the Intended Functionality (SOTIF), tend to start with an identification of potential sources of physical harm and of triggering events that could lead to hazards, and an evaluation of the hazard risk [24]. Though the technical community often conceptualises ethics as an *addition* to safety [35, 63, 80], which is understandable given these established methods, safety is in fact an *ethical* concern. Conceptually speaking, safety is a subset of ethics.[9]

As Peters and colleagues note, however: *'While engineers have always met basic ethical standards concerning safety, security, and functionality, issues to do with justice, bias, addiction, and indirect societal harms were traditionally considered out of scope'* [81]. Meanwhile, the whole field of AI/AS ethics emerged largely because of the risks of different kinds of harm that these novel technologies can incur [82]. Addressing this extended range of harm under the principle of non-maleficence will require both extending the traditional safety envelope and involving people from a range of disciplinary backgrounds.

Depending on the system and its context of use, harms from use of the AI/AS may vary. In addition to the risk of physical harm, there is a risk of psychological harm, such as addiction, anxiety, or trauma [80]; invasions of privacy and misuse of personal data [22, 23, 83]; and harms consequent upon data-driven systems whose use perpetuates discriminatory bias [18-21].[10] Other harms – and harms may be intersectional – might be economic, such as loss of employment from the automation of jobs [84], or crowd-sourced cheap labour to support AI development, e.g., for content moderation and for tagging images for computer vision systems [85]. There is also the risk of societal harms, whether social fragmentation from hyper-personalised algorithms [82], widespread surveillance [86, 87], or damage to essential infrastructure [88]. Further, an AI/AS may directly cause environmental damage during operation, or it may be the product of environmentally unsustainable development and construction, such as from the carbon footprint from training its ML-models or from the use of rare earth metals to build it [31, 85, 89, 90].

---

[9] To clarify, safety concerns are ethical concerns (because avoiding harm is an ethical concern); that is *not* to say that all ethical concerns are safety concerns (because avoiding harm is one of *several* ethical concerns).

The *principle of respect for human autonomy* requires that an important right of human beings – that they should control, "to some degree, their own destiny" [91: 369] – is respected and protected as the power and influence of these advanced technologies, and those who develop and produce them, increases. Human autonomy may be constrained by AI/AS in manifold ways [92]. For example, an AI/AS might unduly 'nudge' people into actions they would not otherwise perform [93, 94]. Or it might be a conduit of misinformation that decreases users' capacity to direct their lives rationally [95]. Or the features that inform an AI system's recommendation, e.g., in job-hiring or credit-scoring, might be ones that risk-bearers would not reasonably endorse and have no control over but are subject to – arbitrary features, for example, rather than intrinsic merit [26]. Or an AS might react to different reasons or facts in the world than the human in- or on-the-loop would,[11] and yet that human may be inextricably caught up in its normative consequences [27, 96]. In addition, human autonomy would be constrained if the human-in-the-loop could not physically stop and AI or cyber-physical system or resume manual control when required.[12]

The principle of transparency, which plays a supporting role in the PRAISE framework (the '+1'), requires that there is sufficiently accessible, salient information about both the human decision-making around the AI/AS across the lifecycle (referred to as 'assurance transparency') and what is going on 'under the hood' of the AI/AS, including its output (referred to as 'machine transparency'). This is important given that human intent can sometimes be obscured in complex decision-making systems [26, 97] and given the inscrutability of increasingly complex ML models [17].

---

[10] To note, discrimination is typically considered to be an injustice (since its wrong-making feature is generally explained in terms of an absence of fairness rather than in terms of harm; and not all discrimination necessarily causes harm). But in the PRAISE argument pattern we include discriminatory bias in the argument module that corresponds to the principle of non-maleficence both because it is more practical to address the issue there and because, while discrimination does not always cause harm, it is plausibly always a hazard or possible source of unjustified harm. Questions of fairness are further dealt with in the argument module that corresponds to the principle of justice.

[11] The distinction between in-the-loop and on-the-loop refers here to the distinction between the capacity for intervention (in-the-loop) and oversight (on-the-loop).

[12] Sometimes resuming manual control may not be safe, and this is considered in the argument module that corresponds to the principle of human autonomy (see Figure 6, (AG9)).

## 3.2. Objections to the ethical principles

Despite (or perhaps because of) the proliferation of ethical principles for AI/AS, some strong voices have emerged which caution against them [98, 99]. Since the PRAISE argument pattern is principles-based, we address these objections before presenting the overview of the framework in Section 4 and describing it in Section 5.

Munn, for example, argues that ethical principles are meaningless, isolated, and toothless - and that the turn to AI/AS ethical principles is therefore ineffective [98, 100, 101]. The claim that principles are meaningless is made because they are often abstract, and it is often not explained what contested terms mean, nor how to reconcile conflicts between them. The claim that principles are isolated is made because ethics declarations are often disjoint from real-world industrial practice and engineering training. As Munn puts it, *"Ethics, so lauded in the academy and the research institute, are shrugged off when entering the engineering labs and developer studios where technologies are actually constructed"* [98: 4]. The claim that ethical principles are toothless derives from the lack of mechanisms to enforce compliance [93, 100, 101].

The framework presented in this paper has some immunity against these criticisms. We explain our use of contested terms such as 'autonomy' and 'justice,' as described above. There is a place in the argument pattern to reason explicitly about trade-offs between core values (see Section 5.5). Moreover, the framework does not preclude some of the *"more granular work or even gruntwork"* that Munn endorses as preferable to ethical principles (98: 6), such as digging into the provenance and quality of datasets on which ML-based systems are trained. This would be addressed under the principle of non-maleficence as part of safety engineering concerns, supported by guidance such as AMLAS [102] or model cards [41]. Further, as ethical frameworks increasingly inform regulatory guidance and ultimately approval [29, 37], ethical concerns will be less easily shrugged off by corporations and practitioners.

It should be noted, more specifically, that the four core principles are controversial in academic medical ethics, where they have their origin. One objection is that relying on these four principles will override people's *"proper feeling for the deeper demands of ethics"* [103: 37].

This comes from a casuist perspective: the school of thought that hard ethical cases are unique and should be evaluated by analogy with paradigm cases [104]. But case-based reasoning can be complementary to a four principles framework [105] and such a framework in fact relies on the exercise of judgement [106, 107]. In addition, assurance argument patterns are starting points for systematic thinking, deliberation, and judgement. It is important that the PRAISE argument pattern is framed in this way, and not as a 'checklist' to be worked through uncritically.

Principles alone will not guarantee ethical AI/AS [73]. Ethical principles are not *sufficient* to ensure positive outcomes. The claim in this paper is not even that the '4+1' ethical principles are *necessary* to achieve ethically acceptable uses of AI/AS. Other approaches could do the job. Rather, the claim is that, suitably worked into a practicable framework, ethical principles in general – and the '4+1' ethical principles in particular – offer a *credible* and *promising* contribution to the achievement and assurance of ethical AI/AS.

## 4. OVERVIEW OF THE PRINCIPLES-BASED ETHICS ASSURANCE ARGUMENT PATTERN

Having introduced the assurance case methodology and the '4+1' ethical principles, let us now bring them together and present the principles-based ethical assurance – or PRAISE – argument pattern. The modular form of the argument pattern is given in Figure 2.[13] Modules are described in the legend in Figure 1. The purpose is to show the flow of the argument pattern before describing its components in more detail in Section 5.

---

[13] Modularity was introduced into GSN by Kelly, in 2001, to support a compositional approach to reasoning about complex systems, initially Integrated Modular Avionics [108, 109].

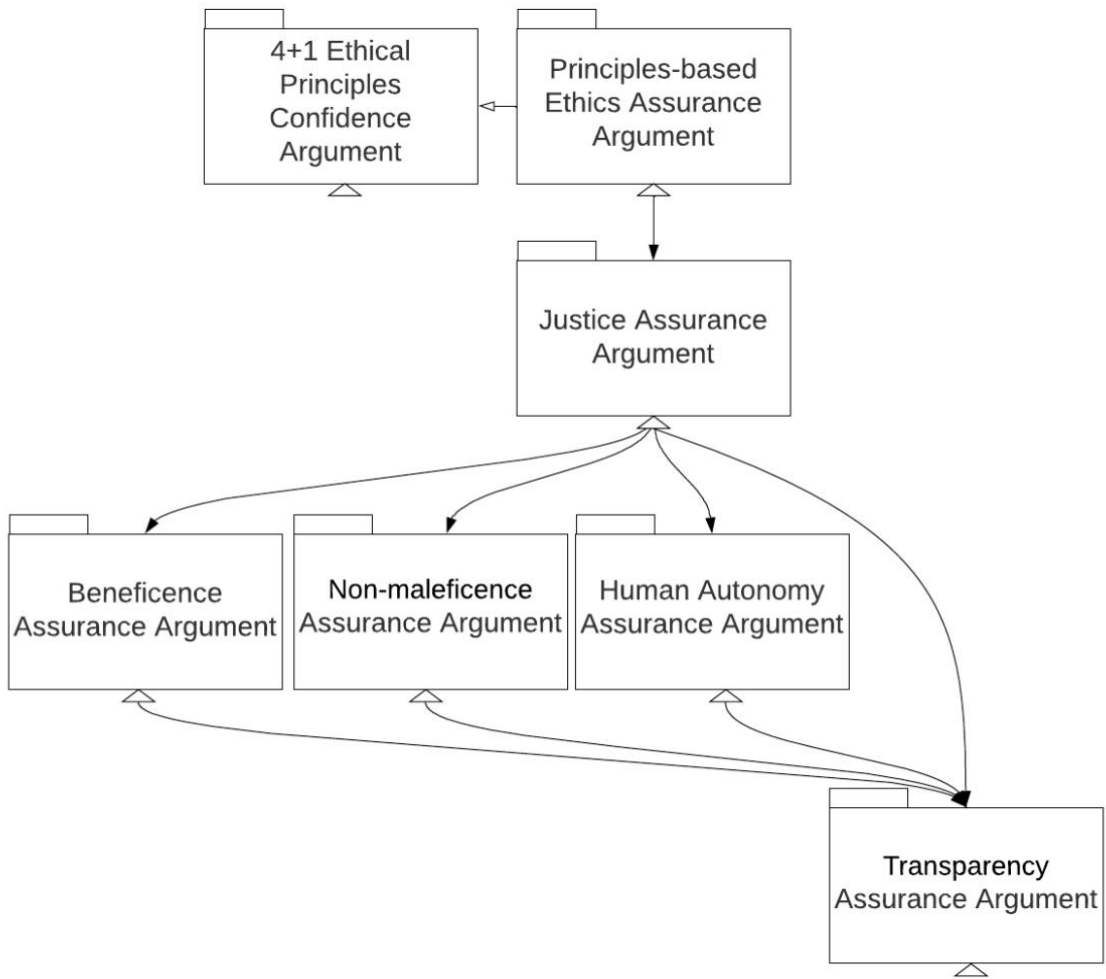

**Figure 2**
**Modular structure of the PRAISE argument pattern**

The module at the top of Figure 2 – the Principles-based Ethics Assurance Argument module – contains the highest-level goal of the argument pattern. The goal of that module – the key claim that the module, and hence the argument pattern as a whole supports – is that, *for the intended purpose, the use of the AI/AS will be ethically acceptable within the intended context*. The entire argument structure is intended to provide justified confidence in the truth of the claim expressed in that goal.

The goal is ethically acceptable *use* of the AI/AS rather than ethically acceptable *design and development* on the grounds that *use* is of the highest generality and perhaps the greatest value. Even if users have good intentions and use the system well, the use of a system cannot be acceptably ethical if its design and development has been unethical. Meanwhile, a system may be designed and engineered impeccably but be misused. Focusing on use covers both cases.

Immediate support for the highest-level goal comes from the Justice Assurance Argument (hereafter, the Justice Argument). The Justice Argument has the sub-goal that the distribution of benefits, tolerable residual risk, and tolerable constraints on human autonomy is equitable across all affected stakeholders. The idea is that, if this distribution were equitable, no affected stakeholders could reasonably reject the decision to deploy the AI/AS, and hence – on the social contract notion of 'ethically acceptable' employed – use of the AI/AS would be ethically acceptable.

The Justice Argument is in turn supported by the Beneficence Assurance Argument (hereafter, the Beneficence Argument), the Non-maleficence Assurance Argument (hereafter, the Non-maleficence Argument), and the Human Autonomy Assurance Argument (hereafter, the Human Autonomy Argument). These three arguments contain sub-goals about actualising benefits, managing risks of unjustified harm, and addressing undue constraints on human autonomy, respectively. Instantiation of each of these argument modules also yields matrices - a benefits matrix, a residual risks matrix, and human autonomy matrix – which provide the information for reasoning about equitable distributions in the Justice Argument.

The role of the Transparency Assurance Argument (hereafter, the Transparency Argument) is to support the other argument modules. Its aim is to provide the information necessary – of the right quantity, quality, relevance and in the right manner – to establish confidence in the claims made across the argument.

At the top of Figure 2, adjacent to the Principles-based Ethics Assurance Argument module, is the 4+1 Ethical Principles Confidence Argument module. This module is presented in Appendix 1 to this paper; it summarises the discussion in Section 3, with the intention of providing confidence in the claim that the 4+1 ethical principles provide a plausible normative basis to achieve ethically acceptable uses of AI/AS.[14]

## 5. DETAILED VIEW OF THE PRINCIPLES-BASED ETHICS ASSURANCE ARGUMENT PATTERN

We now proceed to 'open' the modules of the PRAISE argument pattern. This provides a more detailed view of the proposed framework, but the discussion still proceeds at a relatively high level of abstraction. The imagined example of an autonomous 'robo-taxi' service provides an indication of how the PRAISE argument pattern might apply to individual use cases. Real-world individual ethics assurance cases based on this template would be instantiated in more granularity and with a greater degree of detail. It should also be emphasized that the purpose here is to present and describe the PRAISE argument pattern as a conceptual model; we do not describe how the analysis at each stage of the argument would be conducted.

---

[14] The idea of separating out the main argument in traditional safety assurance cases and the confidence argument was introduced by Hawkins and colleagues [110] to offer a more expressive and more sustained justification for a particular part of the argument than the use of the justification symbol and element in GSN (see Figure 1).

The modules - or individual sub-arguments - are presented in the Goal Structuring Notation (GSN). The symbols and elements in the figures below are consistent with the legend in which these were introduced in Figure 1.

Additionally, for ease of interpreting Figures 3-8 here in Section 5, usage of GSN is explained as follows. *Goals* are labelled with a 'G', e.g., the highest-level goal of the PRAISE argument pattern is labelled 'HG1' in Figure 3. *Contextual artefacts*, such as the definitions, descriptions, or identifications for key terms with the framework, are labelled with a 'C', e.g., the description of the principle of beneficence is labelled 'BC1' in Figure 4. *Argument strategies*, which elucidate the inference between a goal and other goals which support it, are labelled with an 'A', e.g., the strategy of providing justified confidence in the goal of the Non-maleficence Argument by demonstrating that hazards as sources of unjustified harm have been managed is labelled 'NA1' in Figure 5. *Away goals*, which refer to a goal in a different argument module, have a folder icon beneath them. In Figures 4-6, which are the GSN diagrams for the Beneficence, non-Maleficence, and Human Autonomy Arguments, *multiple instantiations of the same goal* are denoted by a solid circle.

## 5.1. The highest-level goal of the argument pattern

Figure 3 below details the highest-level goal of the PRAISE argument pattern, and its immediately supporting elements. This is the goal that the argument pattern, as a whole, is designed to achieve.[15]

---

[15] To be clear, as per the legend in Figure 1, goals in GSN contain claims, and the purpose of the assurance argument is to support justified confidence in those claims or, more specifically, to provide defeasible reasons for believing in the truth of those claims. For the sake of simplicity, however, we refer to them for the most part as goals rather than claims.

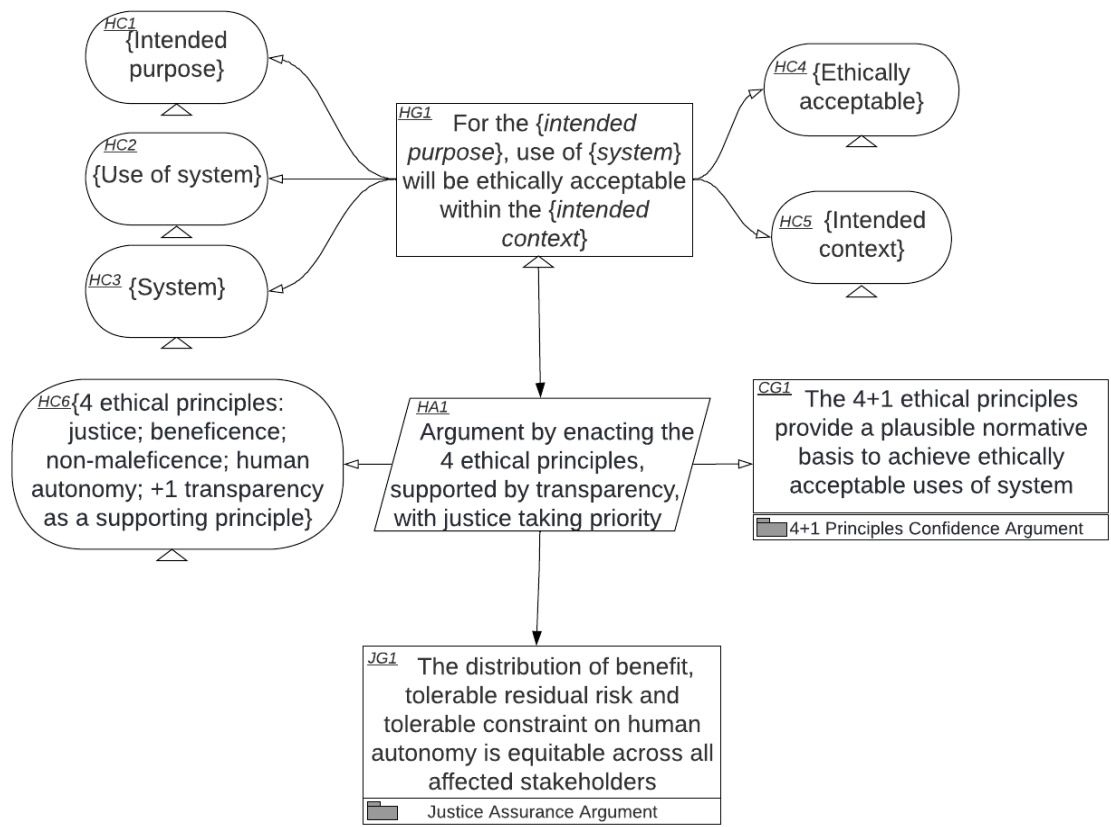

**Figure 3**

**Highest-level goal of the PRAISE argument pattern with immediately supporting argument elements**

The highest-level goal (HG1) is that, for the intended purpose, the use of the system will be ethically acceptable within the intended context (HG1). This goal is situated within a series of explicit definitions, called 'contextual artefacts' (HC1-HC5). We employ the following definition of 'ethically acceptable' (HC1):

> *"The use of the AI/AS would be ethically acceptable if affected stakeholders could not reasonably reject the decision to deploy it in the intended context."[16]*

For the rest of the paper, the argument pattern is illustrated with the hypothetical, or imagined, use case of an autonomous 'robo-taxi' service in a city.

The intended purpose (HC2) might be that the robo-taxi should transport passengers within a specified operational design domain (ODD). The usage (HC3) might be that the robo-taxis take passengers from a city's major railway station to various locations within the ODD, which is a 5-mile wide area in the city.[17] The definition of the system (HC4) would be its description, e.g., a cyber-physical system, with a sense-understand-decide-act (or 'SUDA' loop) design pattern, the computational techniques used to build it, and so on, and its particular degree of autonomy, such that it can drive passengers to their destination without the intervention of a human driver. The intended context (HC5) would include details such as: relevant features of the environment and the population; typical traffic and pedestrian flow; and the limits of the ODD.

---

[16] The sense given to 'reasonably reject' is normative rather than descriptive; that is, the aim is not that no stakeholders actually object to the use of the system – because anyone can actually reject or object to any decision – but that it would not be reasonable for affected stakeholders to object to it. What is meant by 'reasonably reject' is discussed further in Section 5.5.

The argument strategy (HA1) is to demonstrate that the '4+1' ethical principles have been enacted in the modular structure set out in Section 4, with the principle of transparency as the supporting '+1' principles, and with the principle of justice taking priority. This strategy requires a further contextual artefact, namely a description of the '4+1' ethical principles (HC6), which was given in Section 3.

The 4+1 Ethical Principles Confidence Argument (CG1) is an 'away goal' because it repeats a claim earlier represented elsewhere, in the 4+1 Ethical Principles Confidence Argument module (see Figure 2, and Appendix 1 for the GSN diagram for this Confidence Argument).

## 5.2  The Beneficence Argument

As outlined in Section 4, the Justice Argument is the module that immediately supports the highest-level goal (HG1) of the PRAISE argument pattern. For ease of exposition, however, we describe the modules in the order in which someone instantiating the PRAISE argument pattern in an individual ethics assurance case would naturally approach it. We start with the benefits from the use of the AI/AS (the Beneficence Argument) here in Section 5.2, then move to a consideration of the risks (the Non-maleficence Argument) in Section 5.3, and then to the constraints use of the AI/AS might pose n human autonomy (the Human Autonomy Argument) in Section 5.4. Then, we consider the equity of the distribution of these three elements across affected stakeholder groups (the Justice Argument) in Section 5.5.

The Beneficence Argument is presented in Figure 4.

---

[17] There will be other facets of the ODD, e.g., the weather conditions in which the robo-taxi can operate, the time of day for providing the service, etc. However, for the purposes of this paper we can focus on the geographical aspects of the ODD.

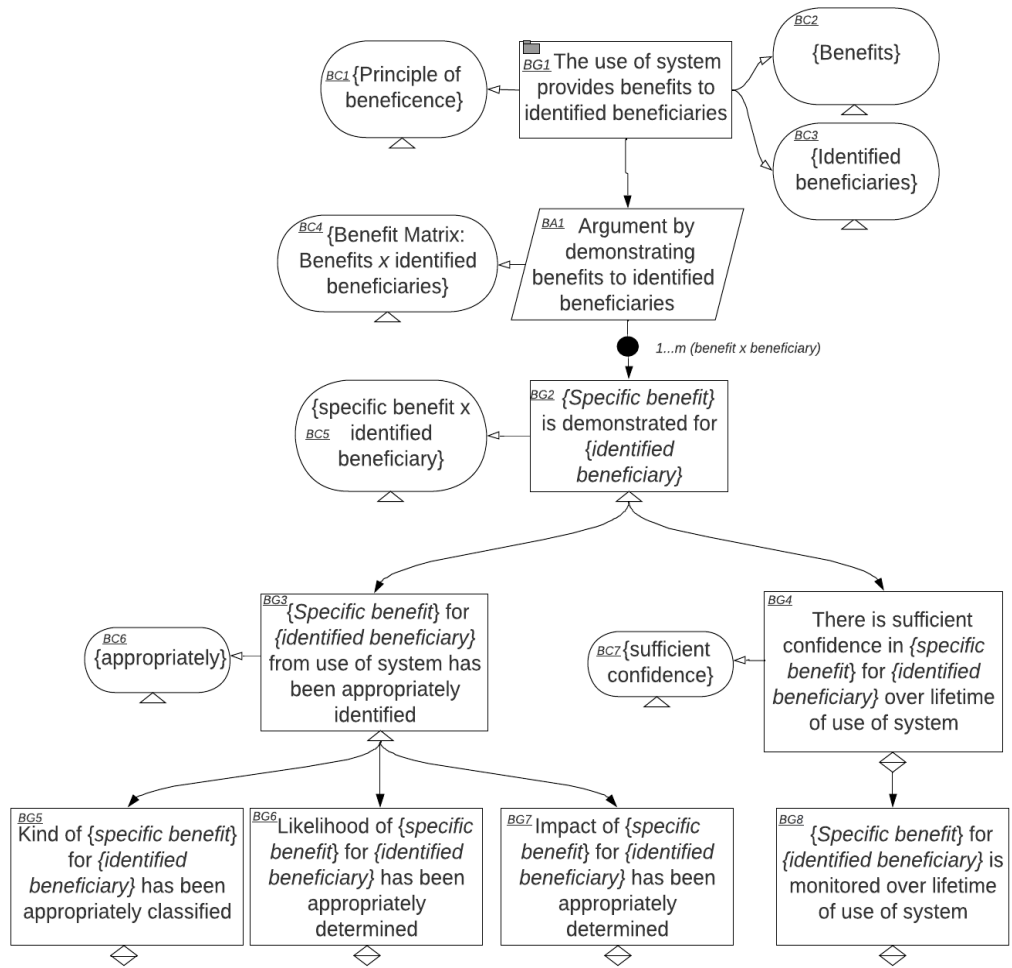

**Figure 4**

**Beneficence Argument module of the PRAISE argument pattern**

The goal of the Beneficence Argument (BG1) is that the use of the AI/AS provides benefits to identified beneficiaries, by which we mean identified beneficiary groups.[18]

These groups of possible beneficiaries are indicated at (BC3). In the illustrative use case of the autonomous robo-taxi service, these might be as follows: i) the end users of the service; ii) other individuals in the ODD, such as pedestrians and local residents; iii) workers in the development and deployment lifecycle of the AVs; iv) the developer or manufacturer of the robo-taxi; v) the service-provider or operator of the robo-taxis; and vi) the municipal or city council.

---

[18] The idea is to aim for a complete picture of the credible benefits and beneficiary groups; it is assumed that instantiators of the argument pattern for individual use cases would work with stakeholders to achieve this.

Having identified the beneficiary groups, the next step is to identify what benefits use of the AI/AS brings to each of these groups. These are indicated at (BC2). To return to the robo-taxi example, benefits from its use might plausibly be as follows. For end users, they might include practical benefits, such as improved mobility and convenience, and possibly financial benefits, such as reduced travel costs. For pedestrians and residents in the ODD, the benefits may be increased physical safety from safe vehicles as well as improved air quality from low carbon, energy-efficient systems [30]. For employees and workers in the development and deployment lifecycle, there may be economic benefits. For service-providers, as well as the robo-taxis' developers and manufacturers, the benefits will likely be primarily financial (with indirect benefits to others in the presence of the right conditions). The benefit to the local municipal council might stem from improved efficiency and reduced costs.

The argument strategy (BA1) is to achieve justified confidence in the goal at (BG1) – in our hypothetical use case, that the use of the robo-taxi service provides benefits to identified beneficiary groups – by demonstrating that these benefits are achieved, or can realistically be expected, for each identified beneficiary group. This is demonstrated by showing that there are specific benefits to these beneficiaries (BG2), which have been appropriately identified (BG3), and about which there is confidence over the lifetime of the use of the system (BG4).

As Figure 4 shows, (BG3) – the sub-goal that each benefit for each identified beneficiary group has been appropriately identified –, which decomposes into three further goals. The first is (BG5), that the *kind of benefit* has been appropriately classified. The suggested approach is to classify these thematically (e.g., safety benefits, financial benefits, environmental benefits). The second is (BG6), that the *likelihood* of the benefit has been appropriately determined, which will require empirical evidence.[19] For example, we would need evidence about the increased mobility for users of the service and the physical safety benefits to pedestrians in the ODD. The third is (BG7), which is that its *impact*, or significance, for that beneficiary group has been appropriately determined.

Measuring the impact of benefits at (BG7) is an open methodological question. It might involve evidence of stakeholder responses, elicited in discussion and consultation. For example, consultations with users of the robo-taxi service, or their representatives, would help to determine how important the benefits of improved mobility, convenience, and reduced travel costs are to this beneficiary group.

---

[19] This may be evidence from testing prior to deployment and should be monitored over the lifecycle, as well as updated if there are material shifts in the context of the AI/AS.

(BG4) – the sub-goal that we can be confident that the identified beneficiary groups will continue to receive each benefit over the lifetime of the system's use – decomposes into the goal that the benefit is monitored over time (BG8), which also requires empirical evidence. For example, physical safety benefits to pedestrians in the ODD, as well as the air quality in the ODD, would require ongoing monitoring.

The role of the Beneficence Argument module is to justify, or provide defeasible reasons for believing, that benefits for identified beneficiaries are actualised or foreseeably actualisable as far as is possible – as Hansson says, *"real risks cannot be traded for hypothetical benefits"* [111: 305]. In addition, the 'benefits matrix' (BC4), which documents these benefits to identified beneficiaries, provides information that is included and reasoned over within the Justice Argument.

### 5.3 The Non-maleficence Argument

The Non-maleficence Argument is presented in Figure 5.

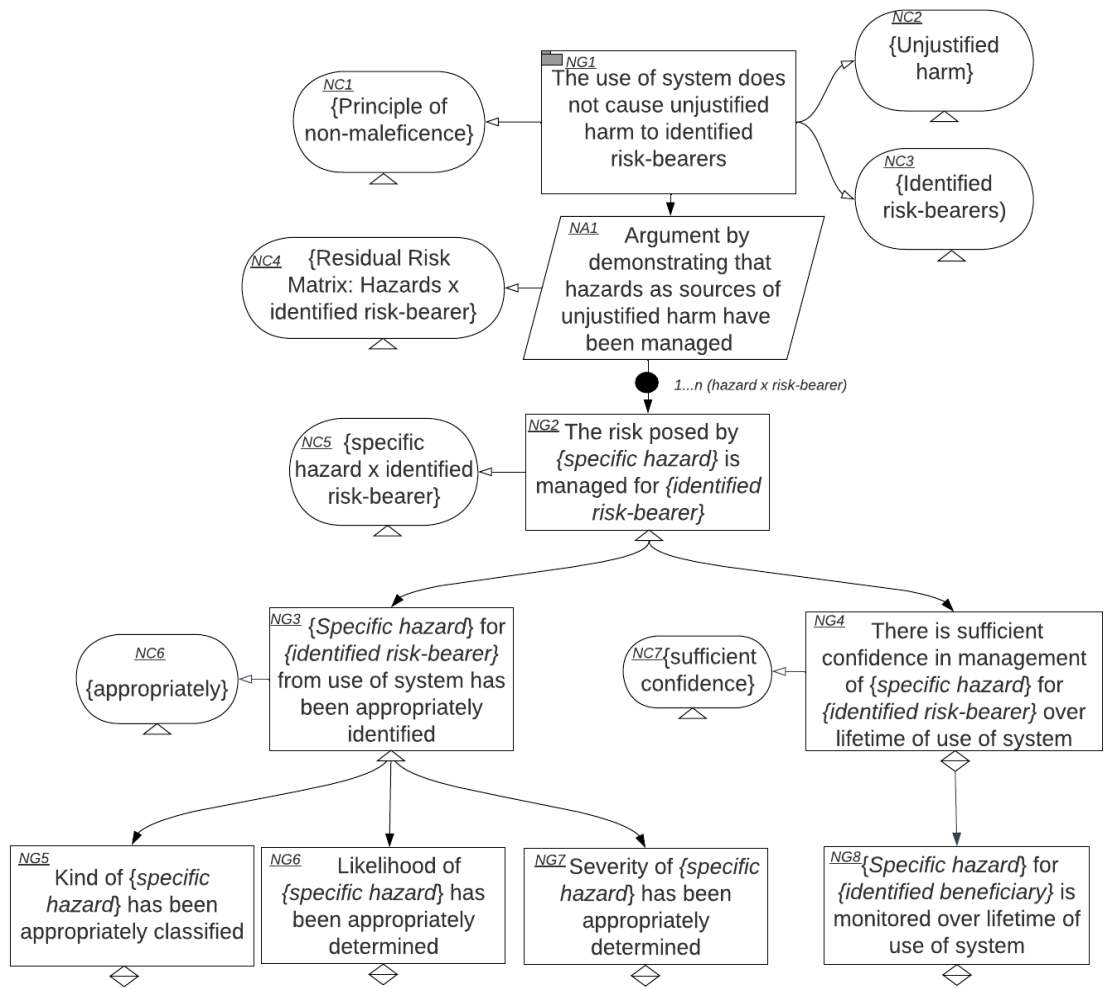

**Figure 5**

**Non-maleficence Argument module of the PRAISE argument pattern**

The goal of the Non-maleficence Argument (NG1) is that the use of the AI/AS does not cause unjustified harm to identified risk-bearing groups.[20] Activities and policies are rarely entirely risk-free, and it is not always *un*justifiable to expose people to *some* risk of harm. It might be justifiable because the residual risks are comparatively small and improbable, or because the risk-bearers benefit from use of the system, or because the benefits in the context are overriding. Take, for example, urgent medical procedures which involve some risk of harm, or the simple act of driving one's car to the shops, which is a permissible risk-raising activity so long as one is careful [111].

---

[20] As with the Beneficence Argument module, the idea is to aim for a complete picture of the credible hazards and risk-bearing groups; it is assumed that instantiators of the argument pattern for individual use cases would work with stakeholders to achieve this.

The identified risk-bearing groups are identified at (NC3). In the hypothetical robo-taxi service use case, these might include: i) end users of the service; ii) other individuals in the ODD, both pedestrians and residents, and local human taxi drivers; iii) workers in the development and deployment lifecycle of the AV; iv) the developer or manufacturer of the robo-taxi; v) the service-provider or operator of the robo-taxis; and vi) the municipal or city council.

Having identified the risk-bearing groups, the next step is to identify the risks of harm use of the AI/AS brings to each of these groups. These are indicated at (NC2). In the robo-taxi example, these might include: physical harm to users or pedestrians; anxiety or stress to users or pedestrians; harms from bias against pedestrians if the ML model which the robo-taxi uses has not been trained on datasets representative of the demographic in the ODD [112]; invasions of privacy, both to users of the robo-taxi and pedestrians captured by the robo-taxi's cameras; economic exclusion; exploitation in the supply chain; environmental hazards across the lifecycle, from the materials used to build it, to the energy-intensiveness of training the ML-models, to the damage it may cause once deployed; financial, legal, and reputational hazards for developers and operators.

The argument strategy (NA1) is to achieve justified confidence in (NG1) – in our illustrative example, that the use of the robo-taxi service does not cause unjustified harm to identified risk-bearers – by demonstrating that specific hazards have been managed for identified risk-bearing groups. This is demonstrated by showing that the hazards have been appropriately identified (NG3) and that there is confidence that the hazards will continue to be managed over the lifetime of the use of the system (NG4).

As Figure 5 shows, (NG3) – the sub-goal that each hazard for each identified risk-bearing group has been appropriately identified –decomposes into three further goals. The first is (NG5), that the *kind of hazard* has been appropriately identified. The suggested approach is to describe these thematically (e.g., physical hazards, psychological hazards). The second is (NG6), that the *likelihood* of the hazard has been appropriately determined, which will require empirical evidence.[21] The third is (NG7), that its *severity* has been appropriately determined. We encounter open research questions about the specification, measurement, and management of non-physical hazards, such as the risk of psychological harm, privacy invasions, discriminatory bias, and economic exclusion – and whether and what the thresholds for tolerable residual risk are for each. These warrant further consideration in the next stage of research.

---

[21] This may be evidence from testing prior to deployment and should be monitored over the lifecycle, as well as updated if there are material shifts in the context of the AI/AS.

Second, there needs to be sufficient confidence that the hazards will be controlled over the lifetime of the system's use (NG4). This further decomposes into the claim that the hazard is monitored over time (BG8), which also requires empirical evidence. For example, that the robo-taxi in operation not only does not cause physical injury in the ODD, but also does not invade the privacy of pedestrians and residents in the ODD, would need to be monitored over time.

The role of the Non-maleficence Argument is to justify, or provide defeasible reasons for believing, that risks of unjustified harm to identified risk-bearers are managed as far as possible. In addition, the 'residual risk matrix' (NC4), which documents the managed and tolerable residual risks to identified risk-bearers provides information that is included and reasoned over in the Justice Argument.

## 5.4 The Human Autonomy Argument

The Human Autonomy Argument is presented in Figure 6.

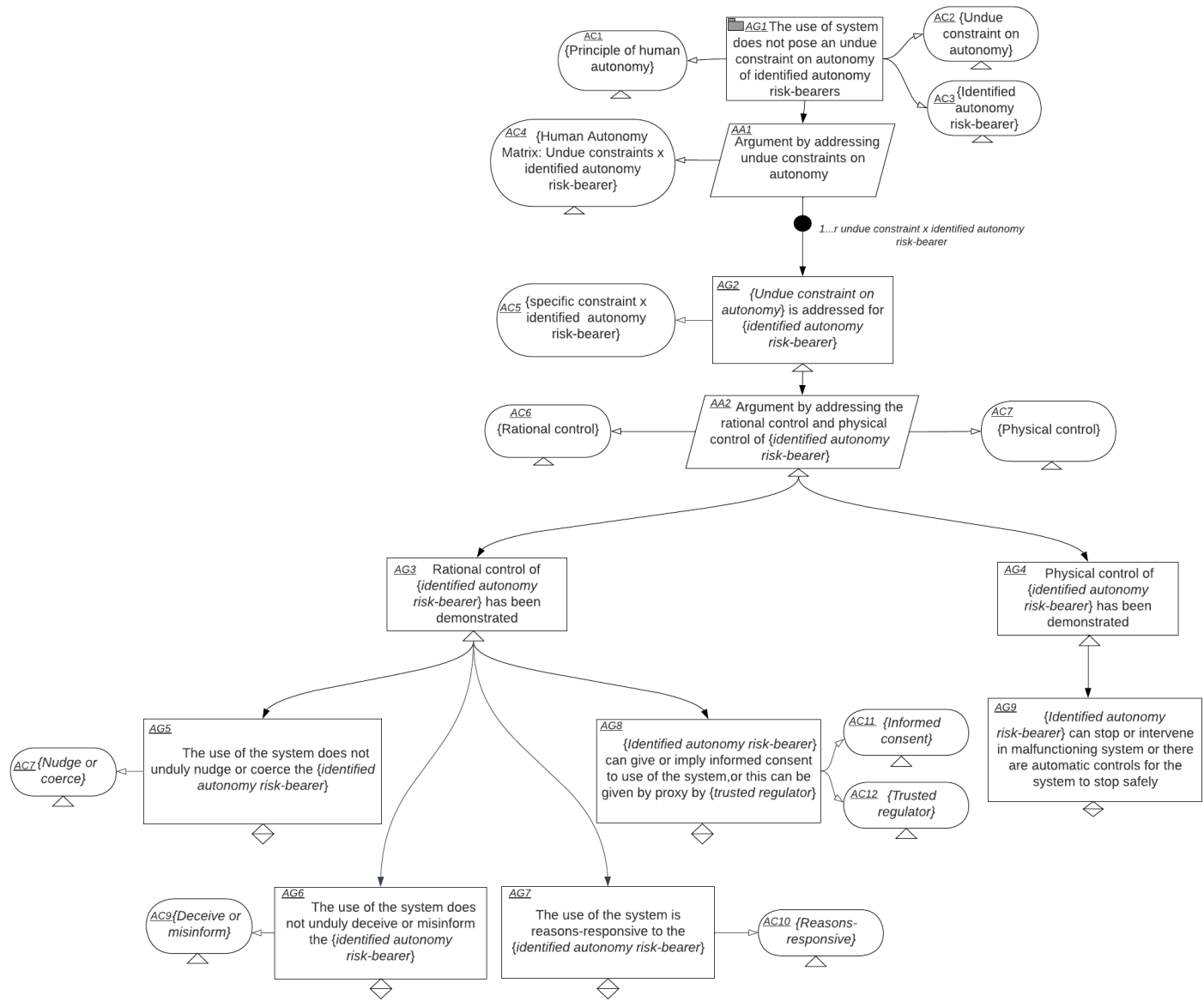

**Figure 6**
**Human Autonomy Argument module of PRAISE argument pattern**

The goal of the Human Autonomy argument (AG1) is that the use of the AI/AS does not pose an undue constraint on the autonomy of identified 'autonomy risk-bearers'.

We take the 'autonomy risk-bearers' (AC3) to be a subset of the risk-bearers in the Non-maleficence argument. They are those classes of individuals in the immediate vicinity of the AI/AS, principally downstream, but also possibly workers upstream, too, whose autonomy is impacted during the production of the AI/AS. In the hypothetical robo-taxi use case, these could be the users of the service, other individuals in the ODD, including other road-users (including taxi-drivers), pedestrians, and residents), and also directly affected workers in the development and deployment lifecycle.

What constitutes an *undue* (as opposed to justified) constraint on human autonomy (AC2) remains a question for debate, but the *kinds* of constraint that would be given at (AC2) are those that are identified in sub-goals (AG5-AG8). After an initial decomposition, which proceeds through an argument strategy that addresses each constraint for each autonomy risk-bearing group (AA1 and AG2), the argument is structured by way of a strategy at (AA2) which organises the constraints upon human autonomy into two categories: constraints to the autonomy risk-bearers' *rational control* over the AI/AS and constraint to their *physical control* over the AI/AS.

Sub-goals (AG5-AG8) address rational control. (AG5) identifies the need to ensure that the use of the AI/AS does not unduly nudge or coerce autonomy risk-bearers. This is to protect these stakeholders' capacities to form their own well-reasoned preferences. (AG6) identifies the need to ensure that use of the AI/AS does not deceive or misinform. This is to ensure that it does not undermine the autonomy risk-bearers' capacities to form their own true beliefs. (AG7) is the claim that the AI/AS is appropriately reasons-responsive to the autonomy risk-bearers. By 'reasons-responsive' (AC10), we mean that the features of the world that the AI/AS responds to are those that the autonomy risk-bearers would endorse.[22] For example, a lone passenger in the robo-taxi at night might not want the vehicle to collect other passengers, such as a drunk stag party.[23] (AG8) covers the idea that autonomy risk-bearers should be able to give their informed consent (AC11); that is, that they should have both the chance and relevant information to agree to the use of the AI/AS, or to opt-out. Where it is not feasible for autonomy risk-bearers to give this, it seems acceptable that it is given by trusted (and trust*worthy*) regulators on their behalf. Substantiating claims (AG7) and (AG8) necessitates appropriate engagement with autonomy risk-bearers during concept design, to elicit what features of the ODD they deem salient to the system's decision-making function, as well pre-deployment, to identify what information supports informed consent.

---

[22] This is an adaptation of the reasons-responsiveness strand of the philosophical literature on moral responsibility, which explains responsibility-grounding autonomy in terms of an agent's sensitivity to reasons. See, for example, [113, 114].

[23] There may be conflict between those features of the world that different groups of autonomy risk-bearers take to be morally relevant. While this is a question for further work, a guiding assumption is that these should be well-reasoned as far as possible.

Sub-goal (AG9) addresses physical control. Discussion of 'meaningful human control' over AI/AS tend to focus on the sorts of concerns addressed by (AG7) and also the capacity, which should be one that can be exercised effectively in practice, to stop or intervene in a system that is going awry [27]. It should be possible, for example, for relevant individuals in the ODD to be able to physically stop or avoid the robo-taxi if it starts to malfunction. Sub-goals to support (AG9) are not given in Figure 6 but would include the requirements that autonomy risk-bearers' have sufficient time, knowledge, and skills to intervene effectively, and the controls are sufficiently accessible. In some cases, where manual override is not possible, there should be automatic controls in place for the system to stop safely.[24]

As with the Beneficence Argument and Non-maleficence Argument arguments, the role of the Human Autonomy argument is to justify, or provide defeasible reasons for believing, ensure that undue constraints on human autonomy (for the subset of risk-bearers who qualify as 'autonomy risk-bearers') are addressed as far as possible. In addition, the 'human autonomy matrix' (AC4), which documents the addressed constraints on the autonomy of identify autonomy risk-bearers, provides information that is included and reasoned over in the Justice Argument.

## 5.5. The Justice Assurance Argument

Figure 7 presents the decomposition of the Justice Argument.

---

[24] Indeed, in the case of the robo-taxi, it may be preferable for the system to stop safely and then to hand control to the occupants or a remote operator, to prevent the occupants becoming 'moral crumple zones' – being held responsible for things over which they have limited control – if an accident occurs [114].

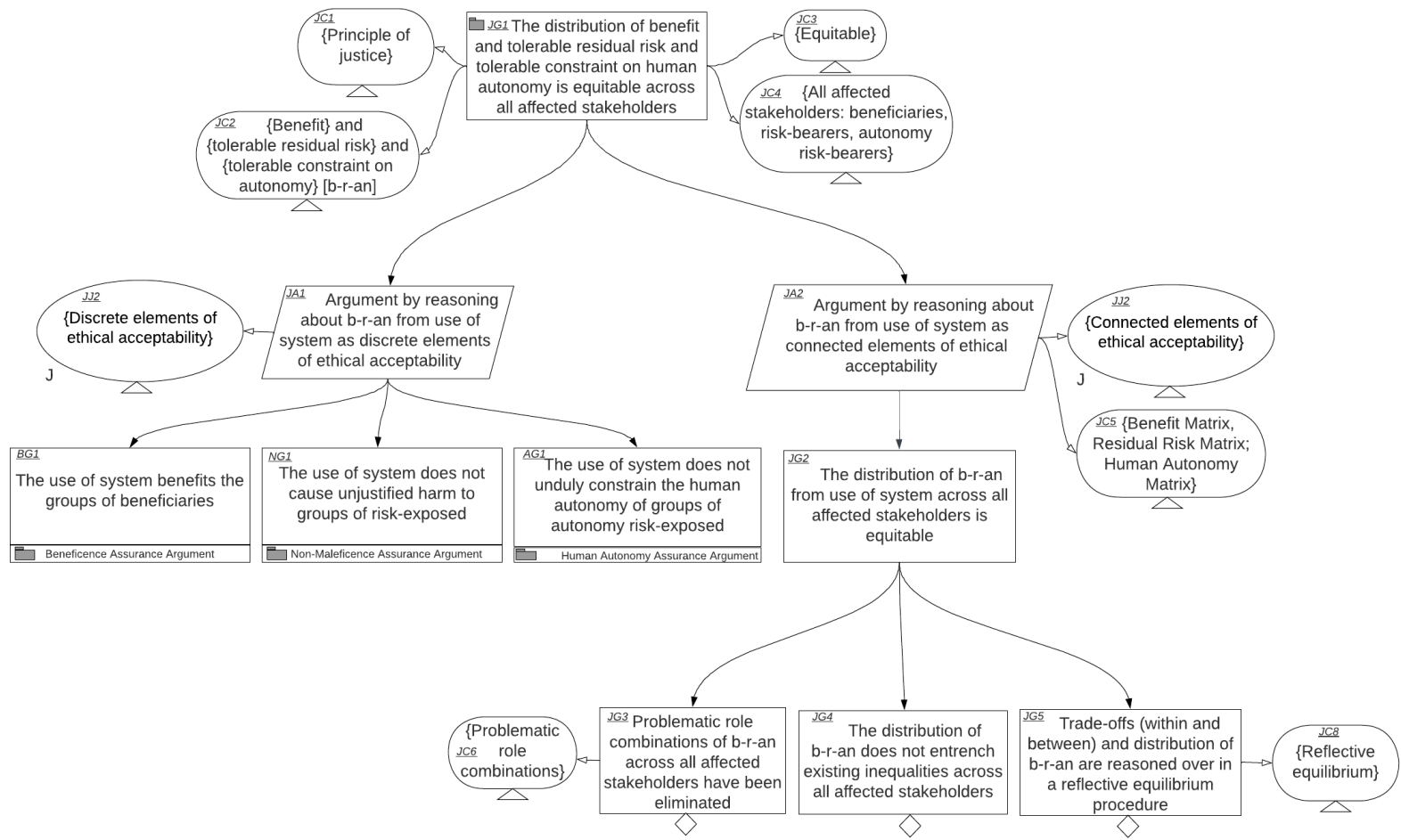

**Figure 7**
**Justice Argument module of the PRAISE argument pattern**

The goal of the Justice Argument (JG1) is that the distribution of benefit, tolerable residual risk, and tolerable constraint on human autonomy (from use of the AI/AS) is equitable across all affected stakeholders. By 'equitable') (JC3), we mean that the distribution is one which recognises existing structures of injustice and asymmetries between different stakeholder groups and aims to treat people fairly in the light of them [111, 115, 116].

The Justice Argument is split into two argument strategies (JA1 and JA2). The argument strategy at (JA1) is to reason about benefit, risk, and autonomy as discrete or separate elements of ethical acceptability. This exports up the reasoning within the Beneficence, Non-maleficence, and Human Autonomy Argument, modules (represented by the away goals BG1, NG1, and AG1) about the provision of benefits, the management of hazards (including, but not limited to, safety hazards), and the constraints on human autonomy that have been addressed.

The argument strategy at (JA2) is to take the three matrices (JC5) from the Beneficence, Non-maleficence, and Human Autonomy Argument modules and use these informational elements to reason about the distribution of benefits, tolerable residual risks, and tolerable constraints on human autonomy across affected stakeholders as *connected* elements of ethical acceptability. As shown in the legend in Figure 1, an argument strategy in GSN elucidates an inference that exists between a goal and its supporting goals. The supporting goal under (JA2) is that the distribution of benefits, tolerable residual risks, and tolerable constraints on human autonomy is equitable across all affected stakeholders (JG2).

We now proceed, in subsections 5.3-5.5, to the decomposition of the Beneficence, Non-maleficence and Human Autonomy Arguments as they would be exported to the Justice Argument in the away goals BG1, NG1, and AG1 under (JA1). In subsection 5.6, we discuss the section of the Justice Argument that falls under (JA2).

To achieve this sub-goal (JG2), three further goals must be achieved: (JG3); (JG4); and (JG5). This is where trade-offs are considered.

In (JG3), ethically problematic role combinations are eliminated. approach is inspired by the ethical risk analysis work of Hansson [115], who describes 'problematic role combinations.' The aim it to eliminate ethically impermissible role combinations of beneficiary, risk-bearer, and autonomy risk-bearers.

Let us elaborate on what these ethically problematic role combinations could be. First, there should be no stakeholder group which *only bears risk* from use of the AI/AS. For example, it should not be the case that pedestrians in the ODD only bear risk from the autonomous robo-taxi service, with no benefit to themselves, and it should not be the case that human taxi-drivers potentially facing job replacement receive no benefits or meaningful compensation. Second, no risk-bearers who receive *only minimal benefit* should be left uncompensated. For example, pedestrians in the ODD might receive only minimal safety benefits from use of the robo-taxi service and no other benefits; in such a case, the onus would be on the relevant decision-makers to increase these benefits and/or to provide others in compensation. Third, no risk-bearer should have their *autonomy unduly constrained*. Reasoning and activities in the Human Autonomy Argument module should already have ensured that there are no severe undue constraints upon the autonomy risk-bearers. However, where it is unclear what constitutes 'undue', this question can be re-considered at this stage of the PRAISE argument pattern. The elimination of problematic role combinations at (JG3) may require adjustments to AI/AS design, or to its intended use or integration into the operating environment, or to the design of its operating environment. If problematic role combinations cannot be eliminated, our working assumption is that it would not be ethically acceptable to deploy an AI/AS as described in (TC1-TC5).

Once problematic role combinations have been removed, the sub-goal (JG4) is to ensure that any existing inequalities are not *entrenched* by use of the AI/AS. The purpose of (JG4) is to consider whether inequalities are reinforced, given that these are technologies which may well *"operate on people and spaces that are already economically and socially stratified"* [98: 1]. Here, trade-offs between fairness and the provision of benefit, the management of risk, and the control of constraints on human autonomy are considered, respectively. For example, in our autonomous robo-taxi example, it may be the case that a subgroup of users do not benefit from the increased convenience of the service if it is not made sufficiently accessible to users with disabilities, even though the wider cohort of users benefit overall. Or increased convenience for users could shift the burden onto low-income residents in the ODD who may have to go further to access affordable travel as a consequence of the service. Substantiating (JG4) will require careful consideration of the demographic, social, and economic context in which the AI/AS is to be deployed, and the inequalities that exist in these spaces. Again, this reasoning may necessitate adjustments to AI/AS design, or to its intended use or integration into the operating environment, or to the design of its operating environment.

The final step of the Justice Argument module occurs at sub-goal (JG5). Here, the intent is that a multidisciplinary team and affected stakeholders (or their trusted representatives) reason about and reach agreement on the distribution of benefits, tolerable residual risks, and tolerable constraints on human autonomy that have 'survived' the earlier stages of the PRAISE argument.

We propose a decision-procedure known as the 'Reflective Equilibrium' (JC8) for this reasoning. Most closely associated with the philosopher John Rawls as methodology for deriving the very principles of justice [117, 118],it is adapted here as decision-procedure about a more specific question: trade-offs between and the distribution of benefit, tolerable residual risks, and tolerable constraints on human autonomy across affected parties from the use of an AI/AS.

The proposed procedure involves a multidisciplinary team and affected stakeholder groups (or their trusted representatives) considering the three matrices (JC5), incorporating adjustments made at (JG3) and (JG4). The Reflective Equilibrium procedure is to work back and forth between: a) their intuitive judgements about areas of concern; b) ethical principles (which may not be exclusive to the '4+1' ethical principles of the PRAISE argument pattern as a whole); and c) relevant non-ethical judgements (e.g., technical, financial, legal) that should be factored into the decision – and then to consider together and propose realistic adjustments that would make the distribution more ethically acceptable. Reflective equilibrium is reached when none of the parties involved are required to make unpalatable compromises and further adjustments in order to endorse the distribution of benefit, tolerable residual risk, and tolerable constraint on human autonomy across affected stakeholders from use of the AI/AS.

As a very high-level illustration involving the robo-taxi example, consensus on the distribution might be conditional on the following adjustments: that the environmental benefits from the use of the service are closely monitored, with particular reference to the air quality experienced by residents in the ODD, and that the local council agrees to make this data public and re-evaluate the use of the robo-taxi should this fall beneath an agreed threshold; that concrete modifications are made to the design of the ODD for vulnerable user and pedestrian subgroups; that residents in the ODD are further consulted to ensure that their daily behaviour is not unduly nudged from the operation of the robo-taxi service; that sufficient guarantees are given from both developers and the municipal operator of the service that user privacy will be protected; and that there is adequate compensation for the human taxi drivers, e.g., through re-training or re-deployment.

It is important that the Reflective Equilibrium procedure at (JG5) is not dominated by self-interested parties with the most powerful voices and the deepest pockets, who may consistently reject a distribution that does not give priority to their own benefit, risk-avoidance, or autonomy. To reiterate, the definition of 'ethically acceptable' used in the PRAISE argument pattern, as set out at (TC4), is that the use of the AI/AS will be ethically acceptable if *no* affected stakeholders could *reasonably reject* the decision to deploy it. Let us now clarify what is meant by 'reasonably reject' – as we signposted in the footnote in Section 5.1. The definition of 'ethically acceptable' derives from a social contract tradition in political and moral philosophy. The central idea of this tradition is that the principles or frameworks we ought to endorse are those which would be reached by hypothetical agreement between rational, autonomous individuals who have equal moral status [119]. This rational agreement or, to adopt T.M. Scanlon's approach [120], a lack of rejection [121], is what provides the justification for the distribution as ethically acceptable. In a situation where the decision about the distribution of benefit, risk, and autonomy was dominated by more powerful players, to the disadvantage of other affected stakeholders, it would be reasonable for those other affected stakeholders to reject that decision.

To avoid this, the question is how to pursue interests in ways that can be justified to others who have their own interests to pursue [119]. The Rawlsian approach would be for the parties participating in the Reflective Equilibrium procedure to imagine themselves behind a 'veil of ignorance', whereby they do not know key facts about who they are or what position in society they occupy [117]. This device helps to ensure that everyone has an equal concern for all. The Scanlonian approach is to assume that individuals are not merely seeking some kind of advantage to themselves but are also aimed at finding a conclusion that others, similarly motivated, could not reject [120]. Both approaches rely on a theoretical fiction, since in the real-world people do not live behind a veil of ignorance and they are not always motivated by moral reasons.

In the context of the argument pattern, it seems less tractable to assume that the individuals deliberating this distribution can behave as if they are wearing a veil of ignorance than it does to assume that they can be guided by and held accountable to a shared aim of ethical acceptability. As such, we recommend that reasoning at (JG4) is carried out by multidisciplinary teams, with adequate representation of the most vulnerable risk-bearing groups, and that an independent party, such as an ethicist, is there to ensure that the aim of the discussion is focused on the goal of reaching an ethically acceptable conclusion which could be justified to all affected stakeholders. Otherwise, the Justice Argument could simply become a forum to prioritise dominant interests, which negates the whole purpose of the framework.

To conclude, working through and achieving sub-goals (JG3), (JG4), and (JG5) would provide justified confidence in, or defeasible reasons for believing in the truth of the claim to, an equitable distribution of benefit, tolerable residual risk, and tolerable constraint on human autonomy across affected stakeholders. Via the argument strategy (JA2), this would provide the necessary support for the goal of the Justice Argument (JG1). Because the structure of the PRAISE argument pattern is such that it is the Justice Argument module that directly supports justified confidence in the claim expressed in the framework's highest-level goal (HG1) – that, for the intended purpose, the use of the AI/AS will be ethically acceptable within its intended context – this is the final connection necessary to complete the PRAISE argument.

## 5.7. The Transparency Assurance Argument

Within the structure of the PRAISE argument pattern, transparency is not included as a core ethical principle in its own right. Rather, as the '+1' ethical principles, it plays a supportive role.

We use the term 'transparency' to refer to the visibility both of human decision-making around the AI/AS across the lifecycle ('assurance transparency') and visibility of what is going on 'under the hood' of the AI/AS, including the explanation of its output ('machine transparency') (TC2 and TC3). This aligns with the definition used within the recently published IEEE P7001 standard on transparency: *"the transfer of information from an autonomous system or its designers to a stakeholder …"* [122: 14]. Less metaphorically, visible information is, at least in part, accessible information. It is also worth pointing out that the existence of an ethics assurance case – whether modelled on the PRAISE argument pattern or on another template – would itself be an exercise in transparency.

Transparency is important just when and because it enables the four core ethical principles to be successfully enacted.[25] It facilitates the transfer of knowledge (i.e., information) to relevant stakeholders, which in turn serves as a basis for them to evaluate confidence in the AI/AS itself and/or in the substantiation of key argument claims. Visibility alone does not guarantee understanding, however. For example, there may be significant differences in stakeholders' epistemic backgrounds that need to be accounted for. This is a consideration for future work.

The model for good transparency or visibility of information is taken from the philosopher Paul Grice's four maxims of cooperative communication [123]. Briefly, these maxims are: quantity;

---

[25] Inappropriate transparency by contrast may work against ethical acceptability. For example, excessive information-giving may distract attention away from a developer's nefarious intentions, and transparency can also be used to undermine a user's autonomy if information about them is used against them [97]. Moreover, transparency is inimical to some security concerns. The equivocal role that transparency can play provides reason not to include it as a separate *core* ethical principle in the framework.

quality; relevance; manner. Ideal rules for participants in a cooperative communicative exchange, they are therefore useful as a standard for assessing the suitability of transparent communication in the context of AI/AS, as others have also highlighted [124, 125].[26]. To ensure that the information provided has the most value for observers and engagers with an ethical assurance case, the Transparency Argument's goals (TG3-TG6) are to ensure that: i) the right amount of information is communicated (quantity); ii) it is truthful (quality); iii) it is salient (relevance); and iv) its transmission facilitates effective exchange of information and understanding (manner).

A high-level version of the Transparency Argument is presented in Figure 8.

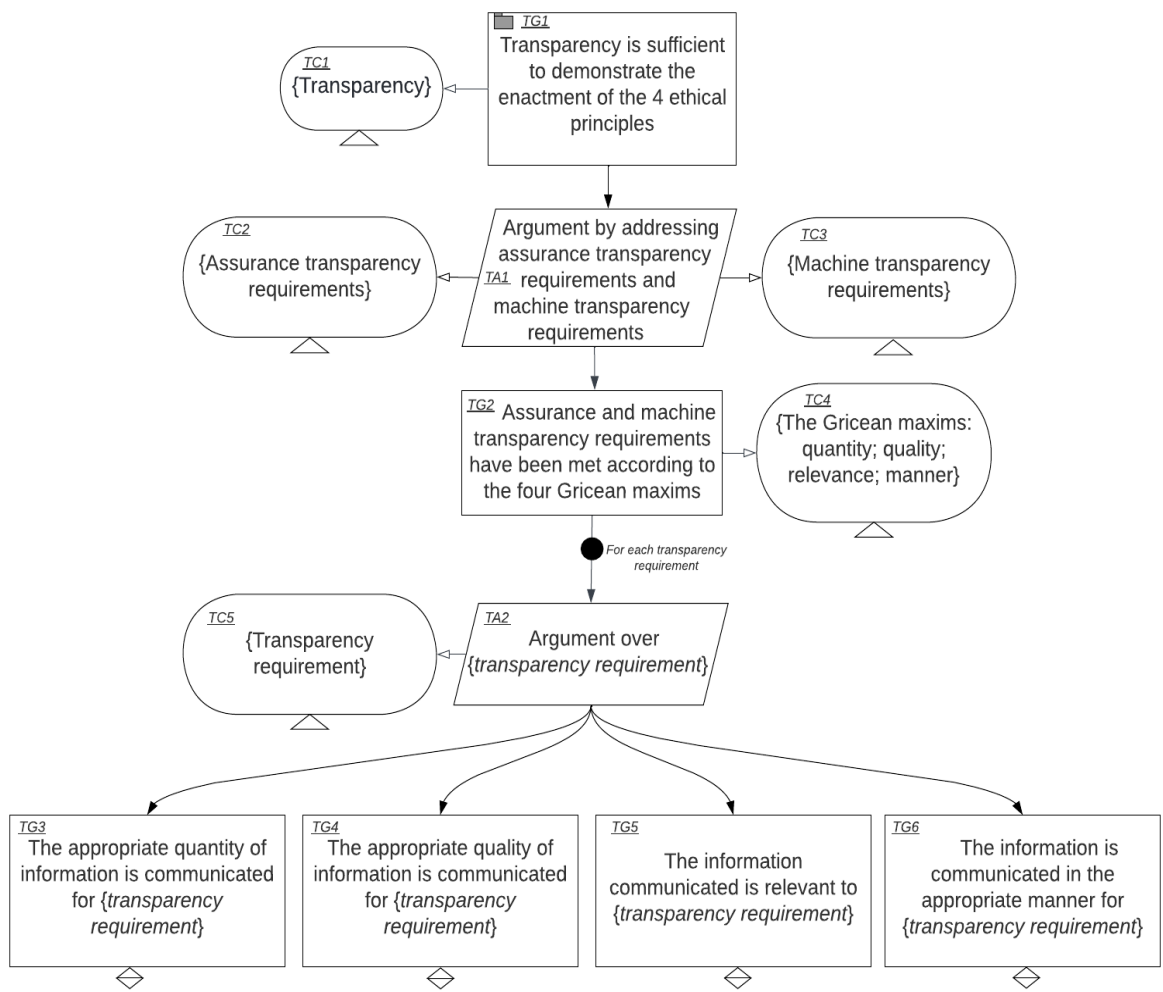

**Figure 8**
**Transparency Argument module of the PRAISE argument pattern**

In discussions in the AI/AS ethics and governance space, 'transparency' is often an elliptical term for machine transparency (TC3). This includes but is not limited to: information about how ML elements are connected to other elements of the AI/AS; dataset auditing; information about the architecture of the AI/AS; and the techniques of Explainable AI. The Gricean maxims offer a standard here. For example, an autonomous robo-taxi may communicate with a remote operator to alert them that it has, or will shortly, exceed its ODD. Alternatively, it may alert a remote operator that adverse weather conditions may prevent its safe operation. In both cases, Explainable AI techniques, such as the use of counterfactual explanations, may facilitate the effective exchange of information (TG6). For instance, the system might communicate to the remote operator that if sensory input does not cease to degrade as a result of inclement weather, control of the vehicle will need to be handed over to a human otherwise it will implement an emergency stop as a safety measure. Machine transparency is therefore necessary, in part, to establish confidence in claims of the Human Autonomy Argument, such as that humans (e.g., remote operators or vehicle occupants) have the appropriate information to take over manual control safely (AG9).

In addition to machine transparency, there is also assurance transparency (TC2) (i.e., roughly, traceability or auditability). This includes but is not limited to: information about why particular decisions were made (e.g., to use one ML technique rather than another, or one dataset rather than another); changelog information; intended usage and ODD information; data storage; and cataloguing information (e.g., for auditing purposes to trace the development of the AI/AS over time). The Gricean maxims offer a standard here, too. For example, a fleet of autonomous robo-taxis may consistently fail to yield to a human operator appropriately in certain locations or under certain conditions. In this case, auditors/investigators may require access to salient information (TG5) about design decisions (e.g., the decision to include or modify certain training data) to assess whether there is a bias (e.g., algorithmic or dataset bias) contributing to the failure to yield. Assurance transparency is therefore necessary, in part, to establish confidence in claims of the Non-maleficence Argument, such as that regulators have the appropriate information to conclude that risks have been managed appropriately (NG2).

Importantly, machine transparency and assurance transparency are not mutually exclusive and can overlap. Indeed, many different aspects of complex AI/AS complicate the distinction between machine transparency and assurance transparency. For example, intentional design decisions to modify training datasets combine elements of machine transparency and assurance transparency. Information about the decision to modify a training dataset (and why) is connected to assurance transparency, whereas the modified training data itself (and access to that data) is connected to machine transparency.

Transparency requirements are diverse and may be intricate. More work needs to be done to establish that pieces of information – whether under the aegis of assurance transparency or of machine transparency – meet the ideals supplied by the Gricean maxims, in particular how the "right amount" of information can be established. The central point here is that the Transparency Argument enables observers and stakeholders to establish confidence that the goals of the PRAISE argument pattern have been achieved.

## 6. DISCUSSION

The principles-based ethics assurance – or PRAISE – argument pattern described in this paper is intended to provide an outline template for individual ethics assurance cases. Engineers, developers, or operators might find value in the framework in that it helps them to structure compelling individual ethics assurance cases which help them to communicate to affected stakeholders the ethical acceptability of the proposed AI/AS in its intended context. It might also benefit regulators as a review or audit tool. Amongst other things, this may help them to answer the 'deployment ethics' question [127] – the question of when the decision to deploy the AI/AS would be ethically justifiable.

The research landscape and range of research issues for ethical AI/AS is vast. The '4+1' ethical principles, with justice as 'first among equals', offer a framework for organising and ordering the contents of this landscape. In this way, it helps to manage complexity. But the PRAISE argument pattern has been presented as a framework at a high level of abstraction. In the speculative example of the autonomous robo-taxi service, for example, we have not distinguished between safety at the level of the vehicle and safety at the traffic level [128], and many other complexities have been overlooked in the interests of presenting the overall conceptual model. The next stage for research is to produce a worked examples of individual ethics assurance cases for real-world AI/AS based on the PRAISE argument template – to evaluate its viability, usefulness, and ability to handle complexity at the more granular level.

Applying the argument pattern to specific use cases will also help us to refine our understanding of – and address – the main practical limitation of the model, which is that specific methodologies for measurement are not yet clearly defined. Instantiating the argument pattern will require the identification of appropriate metrics for several qualitative parameters [31]. But there are open research questions about the measurement of non-financial benefits, risks of non-physical harm, constraints on human autonomy and how the "right amount" of information is established for the Transparency Argument. We may also encounter a scarcity of baseline comparisons before the deployment of AI/AS. It is worth noting, however, that while there are imperatives to address these questions, uncertainty can still be explicitly identified in the three matrices in the argument pattern and factored into judgements during the Reflective Equilibrium procedure about equity and ethical acceptability.

Producing worked examples of individual ethics assurance cases based on the PRAISE argument pattern will also help us to evaluate its effectiveness. The following would be indications that the framework proposed in this paper is effective. First, evidence that it is practicable to instantiate. Second, that it not only provides a covering framework for ethical concerns that are addressed by existing assurance tools, such as standards, but that it also helps participants to identify issues salient to ethical acceptability that would not otherwise come up with other methods. Third, that working through the framework, and particularly the Reflective Equilibrium procedure, leads to concrete, actionable adjustments that improve the equity of impact from use of the AI/AS across affected stakeholders. In the medium to long term, effectiveness of the PRAISE argument pattern would be revealed if and when AI/AS that are subject to individual ethics assurance cases based on the template are more gratefully welcomed, more widely and sustainably adopted, and more positively transformative than systems that are not.

Presently, proposals for regulatory frameworks for trustworthy AI recommend a 'toolbox' of assurance methods, such as technical standards and dataset audits [34, 36]. Ethics assurance cases are starting to be proposed as part of the toolbox for the assurance of AI [58-63]. Most proposed tools cover discrete values, such as fairness and transparency. But questions remain about how to reason over and reconcile competing values and trade-offs. The PRAISE argument pattern offers a framework for this. This something standards in this domain do not currently do – and perhaps, by their very nature, cannot do. Indeed, trade-offs will most often be context-specific and not obviously amenable to standardization. The PRAISE argument pattern could therefore work in tandem with such tools and methods.

It remains important to reiterate two things. First, that the PRAISE argument pattern is not intended to be reduced to algorithms and turned into an automated process. It is intended as a framework for human deliberation and judgement about the ethical acceptability of uses of AI/AS. Second, that it is not intended to be worked by people as a tick-box exercise or checklist. The template is intended to provoke and support, rather than preclude, human reflection and judgement about the ethical acceptability of uses of specific AI/AS in defined contexts.

It is a bold and ambitious framework - not just in considering overall ethical acceptability, but also in having a high threshold of ethical acceptability which is grounded in the notion of a social contract that gives equal respect and status to all affected stakeholders. For this reason, the PRAISE acronym is perhaps apt in that it points to what could be a 'gold standard' for ethics assurance. The underlying belief is that such an approach will help to ensure that the benefits from the development and deployment of AI/AS are reaped by all.

This leads to a specific question for reflection, which is less frequently asked than it should be: who is included in the 'trade space' or scope of ethical concerns When considering the distribution of benefits, risks, and constraints on human autonomy, how far upstream should we go? Should we include all risk-bearers up to the very earliest stages of the supply chain, including the working conditions of those mining the minerals from the ground, or in the factories in which the semiconductors are produced, or where the pixel-precise labelling of images for ML training datasets takes place? By the same token, how far downstream should we go? Clearly, the scope needs to extend to users and those in the immediate operating environment, but what about individuals in ten years' time, or future generations who may be positively or negatively affected by the use of the system? For example, widespread deployment of even beneficial AI/AS may have negative consequences on the environment and impose increased risk on future generations [30]. The temptation is to keep the framework manageable by restricting scope, but the danger is that this becomes a form of wilful myopia and uses of systems that assimilate serious harm pass a threshold of ethical acceptability. If the ultimate motivation for the development and deployment of these technologies is to benefit humanity and to advance equity, as it is often claimed, then the scope of the argument pattern matters radically.

—

APPENDIX 1 - **4+1 ETHICAL PRINCIPLES CONFIDENCE ARGUMENT**

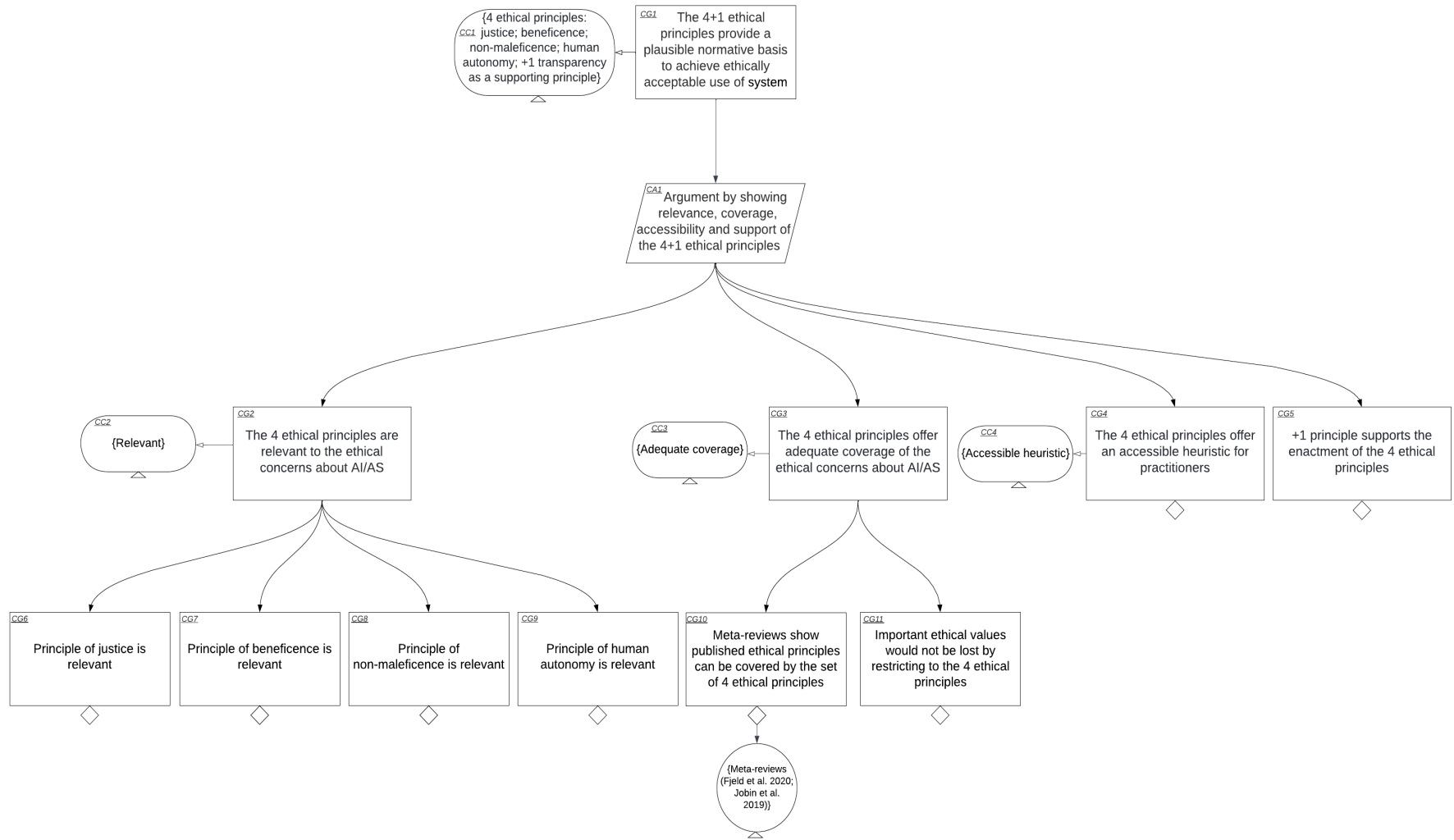

**Figure 9**
**4+1 Ethical Principles Confidence Argument**